# Performance characterization of a novel deep learning-based MR image reconstruction pipeline

R. Marc Lebel, PhD

GE Healthcare, Calgary, Alberta, Canada

## Abstract

A novel deep learning-based magnetic resonance imaging reconstruction pipeline was designed to address fundamental image quality limitations of conventional reconstruction to provide high-resolution, low-noise MR images. This pipeline's truncation aims were to convert truncation artifact into improved image sharpness while jointly denoising images to improve image quality. This new approach, now commercially available as AIR™ Recon DL (GE Healthcare, Waukesha, WI), includes a deep convolutional neural network (CNN) to aid in the reconstruction of raw data, ultimately producing clean, sharp images. Here we describe key features of this pipeline and its CNN, characterize its performance in digital reference objects, phantoms, and in-vivo, and present sample images and protocol optimization strategies that leverage image quality improvement for reduced scan time. This new deep learning-based reconstruction pipeline represents a powerful new tool to increase the diagnostic and operational performance of an MRI scanner.

## Introduction

Magnetic resonance imaging (MRI) provides unique advantages compared to some other medical imaging modalities in that it does not use harmful ionizing radiation, provides excellent soft tissue contrast, and can provide volumetric information. However, MRI is an inherently slow technique, and parameter adjustments to improve image quality typically come at the expense of longer scan times.

In MR imaging, the time domain signal corresponds to the Fourier transform of the transverse magnetization distribution in the object being imaged. Image artifacts may arise as a result of the specific data acquisition and reconstruction processes. Several of these artifacts are ubiquitous, occurring in all images irrespective of the acquisition type. Notably, thermal and electrical noise during data sampling translates into random image noise that reduces the signal-to-noise ratio (SNR) of the image. Incomplete sampling of high spatial frequencies in the Fourier domain (k-space) creates edge ringing, also known as truncation artifact or Gibbs ringing, in the final reconstructed image. Traditional approaches to mitigating these challenges include a combination of hardware, software, and acquisition parameter adjustments. Hardware solutions such as higher field strength magnets, low-noise receive chains, and more radiofrequency coil elements for signal reception have been introduced to improve SNR. The SNR can also be improved by increasing the number of signal averages. Software filters are commonly applied in the data reconstruction pipeline to mitigate noise and ringing; however, these are only partially effective and can have the undesired impact of reducing effective spatial resolution (Figure 1). Truncation artifacts can be mitigated by increasing the acquired spatial resolution, which in turn typically increases scan time while also reducing SNR. Parallel imaging, partial Fourier, and compressed sensing can reconstruct images from sub-sampled data to reduce scan times, but at the expense of SNR and/or spatial resolution. This costly SNR/spatial resolution/scan time interdependency forces clinicians to make difficult tradeoffs in daily practice to sacrifice image quality versus scan time for a given patient and clinical need.

Recently, artificial intelligence (AI)—particularly deep learning (DL)—has re-defined the state-of-the-art in a multitude of tasks including image classification, segmentation, denoising, super-resolution, and image synthesis/transformation (Dong, Loy, He, & Tang, 2014; Gatys, Ecker, & Bethge, 2016; Ronneberger, Fischer, & Brox, 2015; Zhang, Zuo, Chen, Meng, & Zhang, 2017). Deep learning is also being applied to generalized inverse problems (Adler & Öktem, 2017; Schwab, Antholzer, & Haltmeier, 2019), where it can serve as a regularizer to quickly yield plausible solutions. DL-based methods that learn how to reconstruct the image from training on previous data have also shown tremendous potential for a paradigm shift in medical image reconstruction. In MRI, the majority of these methods focus on reconstructing undersampled k-space data to enable scan times that are even shorter than achievable with established acceleration methods (Hammernik et al., 2018; Malkiel et al., 2019; Souza, Lebel, & Frayne, 2019; Zhu, Liu, Cauley, Rosen, & Rosen, 2018). These techniques may potentially enable or improve certain applications that demand the shortest possible temporal footprint or scan time. However, many clinical applications are in greater need of image quality improvement in terms of SNR and spatial resolution, rather than additional acceleration. Many common protocols use only modest or no acceleration rates or even multiple averages for sufficient SNR. An advanced reconstruction method that makes more efficient use of the acquired data to produce high SNR yet detailed images would substantially benefit these acquisitions, which in turn could add more flexibility in protocol adjustments using well established approaches to reduce scan time while maintaining diagnostic image quality.

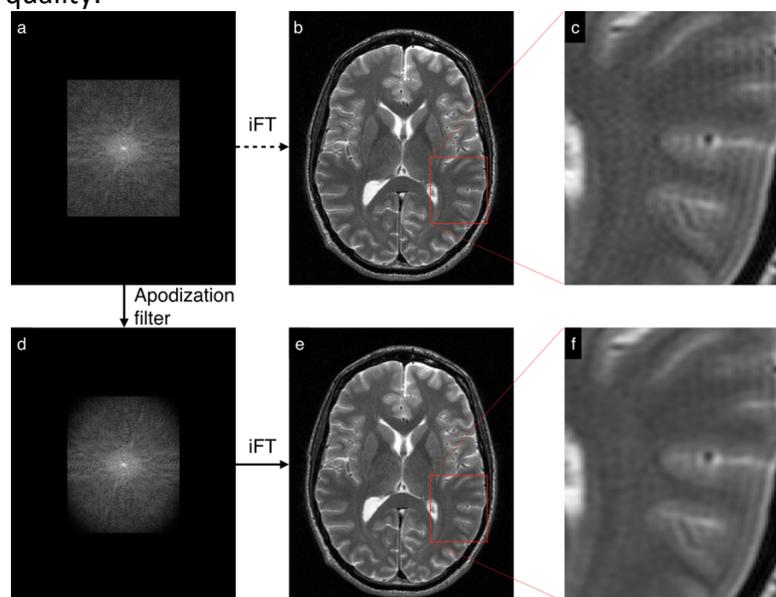

Figure 1: Acquired k-space data (a) can be converted to an image (b, c) via inverse Fourier transform (iFT), but this produces strong ringing and the image has noise throughout. Conventional reconstruction piplines, adopted in some form by all vendors, use k-space (or image space) filters (d) to reduce ringing and noise in the final image (e, f). This linear filtering only partially removes ringing and noise but broadens the point spread function, reducing the image resolution.

In this work we present key technical features, characterize performance, and demonstrate typical uses of AIR™ Recon DL, now available as a product on MR scanners (GE Healthcare, Waukesha, WI), compared to conventional MR reconstruction methods.

## Methods

### AIR™ Recon DL

The AIR™ Recon DL reconstruction pipeline takes raw k-space data as its input and generates high fidelity images as its output (Argentieri et al., 2019; Bash, Thomas, Fund, Lebel, & Tanenbaum, 2019; van der Velde et al., 2019; Villanueva-Meyer et al., 2019). The goal is to produce images that are consistent with the acquired data, have no ringing artifacts and have reduced noise power, ultimately improving diagnostic confidence over conventional methods. AIR™ Recon DL eliminates the need for resolution-degrading filters used in a conventional reconstruction pipeline (Figure 1). The AIR™ Recon DL pipeline includes a deep CNN that operates on raw, complex-valued imaging data to produce a clean output image. Specifically, the CNN is designed to:

- Provide a user tunable reduction in image noise,
- Reduce truncation artifacts, and
- Improve edge sharpness.

Integration into the scanner's native, inline reconstruction pipeline is critical as this provides access to raw, full bit-depth data. This ensures noise is distributed in the complex plane and has predictable and desirable statistics, such as zero-mean. Additionally, both noise and ringing can be represented in domains where they have reasonably finite support, which enables the CNN to operate effectively with a receptive field that is smaller than the data matrix.

The CNN contains 4.4 million trainable parameters in approximately 10,000 kernels. It is a convolutional network, making it suitable for all MR relevant images sizes. The CNN uses no bias terms and employs ReLU activations, which together provide two benefits. First, the network is scale invariant: it is equally applicable to low- and high-intensity images without requiring delicate rescaling. Second, without bias-imposed thresholds and with piecewise linear activations, the network is suitable for effective blind denoising, adapting equally well to any noise amplitude including spatially variable noise within the same image. This network design and its tight integration in the reconstruction pipeline allows the CNN to operate seamlessly with many existing technologies, including parallel imaging (specifically ASSET and ARC) and partial Fourier.

The CNN also accepts a user-specified denoising level, a scalar parameter between 0 and 1 representing the fraction of the estimated noise variance to be removed. The pipeline output consists of an image with all of the estimated truncation artifact removed (which is independent of the denoising level) and the specified fraction of the noise removed. This design preserves image features and enables adjustment based on user preference.

The CNN was trained with a supervised learning approach using pairs of images representing near-perfect and conventional MRI images. The near-perfect training data consisted of high-resolution images with minimal ringing and very low noise levels. The conventional training data were synthesized from near-perfect images using established methods to create lower resolution versions with more truncation artifacts and with higher noise levels. A diverse set of training images spanning a broad range of image content were employed to enable generalizability of the CNN across all anatomies. Image augmentations, including rotations and flips, intensity gradients, phase manipulations, and additional Gaussian noise were applied for added robustness, resulting in a training database of 4 million unique image/augmentation combinations. Training was performed in a single epoch of the training database (i.e., 4 million training iterations). The ADAM optimizer (Kingma & Ba, 2014) was used to minimize the loss between the CNN predicted and the near-perfect images. A generative adversarial network (Goodfellow et al., 2014) was not used to improve image sharpness so as to avoid potential hallucinations of new features (Ledig et al., 2017).

Performance characteristics of the AIR™ Recon DL network are summarized visually in Figure 2 using the large ACR phantom and an axial $T_2$ FLAIR image of the human brain. AIR™ Recon DL images were created with denoising levels of 0.30, 0.75, and 1.0 in this example. Difference images between the AIR™ Recon DL brain images and the original are also presented. The phantom and in-vivo images were chosen to illustrate the broad applicability of the network to different objects and contrasts, insensitivity to image scaling, and adaptability to different noise levels. The CNN effectively eliminates ringing in all cases while denoising is controlled independently of the ringing reduction. The CNN does not alter the appearance of the subtle motion artifacts in the brain image.

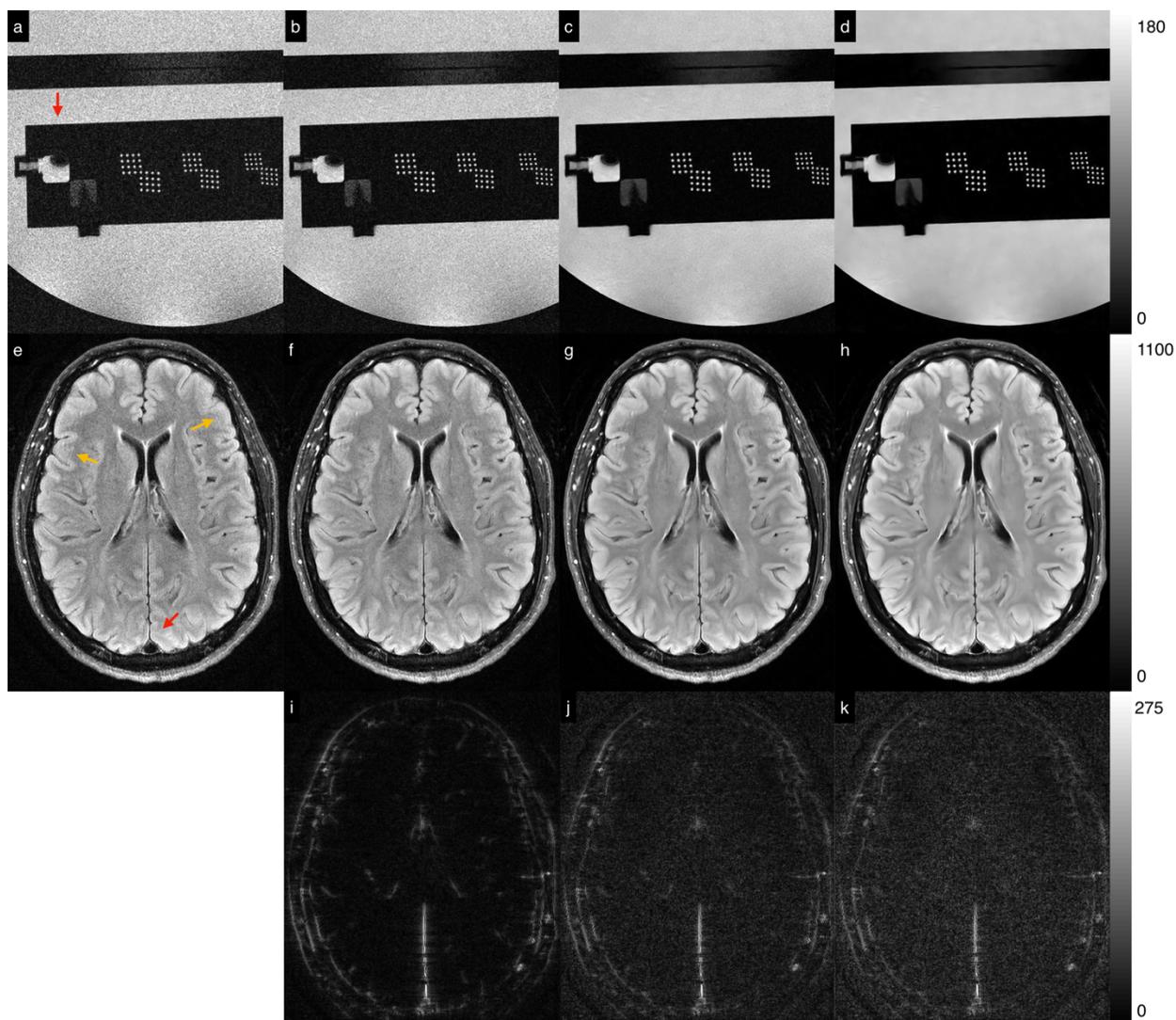

*Figure 2: AIR™ Recon DL performance at different denoising levels in a phantom and in vivo. Original raw images in a cropped portion of the large ACR MRI phantom (a) and in vivo (e). AIR™ Recon DL images at denoising levels of 0.30 (b, f), 0.75 (c, g), and 1.0 (d, h). Difference images (i–k) between the in vivo images reconstructed with and without AIR™ Recon DL show constant ringing suppression but with the specified denoising level. Structures visible at edges in the difference images represent removed truncation artifact, not lost detail. In all cases, AIR™ Recon DL removes ringing from the raw images (red arrows); motion artifacts in the in vivo image (yellow arrows) are unchanged. AIR™ Recon DL is insensitive to scaling of the source images, which differ by approximately an order of magnitude in this example.*

## Performance testing

The performance of AIR™ Recon DL was assessed using two digital reference objects (DRO) for low contrast detectability, phantom measurements of SNR and edge sharpness, and *in vivo* images for qualitative evaluation.

A human observer study was conducted using a multi-object DRO consisting of a grid of circular disks of varying diameter, between 1 and 12 pixels (at the simulated acquisition resolution), and of varying amplitude. Complex white Gaussian noise was added to the object to produce contrast-to-noise ratios (CNR) between 1.0 and 25.0. Forty-eight different noise realizations were evaluated. Data from the noisy objects were input to the CNN along with a denoising level of 0.75. The output images were evaluated by two human readers. For each known disk location, the readers indicated whether the disk was visible or not. A probability of detection, representing the number of times a disk was visible divided by the total number of trials, was computed for the original object and the object output from the AIR™ Recon DL CNN. Readers were not blinded since the image type was obvious but had a cooling off period between the original and the corresponding AIR™ Recon DL image. The difference in detection probability was reported.

A model observer was used to quantify low-contrast detectability in a single-object DRO using a signal known exactly (SKE) task (Tseng, Fan, Kupinski, Sainath, & Hsieh, 2014; Vaishnav, Jung, Popescu, Zeng, & Myers, 2014). The DRO used in this test consisted of noisy images with, and without, a square signal centered in the image. Images were created on a 120x120 grid, a square signal of uniform intensity was added for the signal present cases, and white Gaussian noise was added. Objects of three different sizes (1x1, 2x2, and 4x4 pixels) were evaluated. As described by (Vaishnav et al., 2014), the Hotelling observer's test statistic is given by

$$\lambda(\mathbf{g}) = \mathbf{w}^T \mathbf{g}$$

where $\mathbf{g}$ is a test image (in vector form) and $\mathbf{w}$, a pre-trained template vector. The template is defined as

$$\mathbf{w} = \mathbf{C}^{-1}\left(\bar{\mathbf{g}_1} - \bar{\mathbf{g}_0}\right)$$

where $\mathbf{g}_0$ and $\mathbf{g}_1$ are the average images without, and with, signal present respectively, and $\mathbf{C}$ is the covariance matrix.

Solving the template directly is challenging since inverting the covariance matrix is generally ill-conditioned. With *a priori* knowledge that our signal is sparse and noise is minimally correlated, we expect the template vector to also be sparse. With this knowledge and a very large amount of synthetic data available, the template can be well approximated as:

$$\underset{\mathbf{w}}{\operatorname{argmin}}\left\|\mathbf{Cw} - \left(\bar{\mathbf{g}_1} - \bar{\mathbf{g}_0}\right)\right\|_2^2 + \lambda_r \|\mathbf{w}\|_1$$

where $\lambda_r$ is a mild regularization factor promoting a sparse template. This constrained problem was solved using an alternating direction method of multipliers algorithm (Boyd, Parikh, & Chu, 2011). The regularization factor was set at $\lambda_r = 10^{-4}$ in all cases (for reference, the signal intensity of the rectangular objects is 3.0, 1.5, and 0.75 for the 1x1, 2x2, and 4x4 pixel objects respectively and the noise variance in each image is 1.41). Regularization factors between $10^{-2}$ and $10^{-7}$ produced similar test statistics.

A total of 4096 noise realizations were computed for each signal present/absent condition. These 4096 repetitions were divided into 8 groups (of 512 repetitions): 7 groups were used to compute the template (i.e., train the observer) while the final group was used to compute the test statistic (i.e., test the observer). These groups were permuted to provide a bootstrapped estimate of the mean detectability and uncertainty. Receiver operating characteristic (ROC) curves and area under the curves (AUC) were computed.

The large ACR phantom (ACR, 2018) was used for SNR and sharpness measurements. The resolution insert was scanned with a 2D spoiled gradient echo sequence with a 210 mm field of view, 512x512 matrix, repetition time of 12.6 ms, echo time of 6.0 ms, bandwidth of ±125 kHz, and 3 mm slice thickness. Thirty consecutive images were acquired in a total of 194 s (6.5 s per image). Consecutive images were retrospectively averaged to create a larger dataset consisting of images with multiple averages and repetitions. SNR measurements were made using pairs of images, according to:

$$SNR = S/\left(\sigma/\sqrt{2}\right)$$

where S is the average signal in a homogenous region of interest (ROI) in the first image and $\sigma$ is the standard deviation in the same ROI on the difference between the first and second images. Sharpness was measured by plotting line profiles across abrupt edges in the phantom. Sharpness was quantified as the ratio of the peak slope in the line profiles in the AIR™ Recon DL images relative to the original images. This provided a relative measure of sharpness, independent of absolute intensity. Relative sharpness was measured in four locations and was repeated for images with a different number of averages.

A diverse set of *in vivo* images were acquired in multiple anatomies using a range of clinically relevant pulse sequences and contrast weightings. Images in the spine, wrist, abdomen, heart, and brain were reconstructed using a conventional pipeline and with AIR™ Recon DL at a denoising level of 0.75. The pulse sequences used included fast spin echo (spine, wrist, brain, heart) and single shot fast spin echo (abdomen). Contrast weightings included $T_2$ (spine, abdomen, brain), $T_1$ (spine, brain), and magnetization preparation pulses for fat saturation (spine, wrist) and fluid suppression (brain, heart). A gadolinium-based contrast agent was used in one brain case. Images were acquired at 1.5 T and 3.0 T (GE Healthcare, Waukesha, WI).

The ability to optimize protocols for shorter scan time and slightly higher resolution was demonstrated in two common brain scans at 3.0 T. The scan time of a $T_2$ weighted acquisition was reduced from 127 to 61 sec by adding a 1.5-fold parallel imaging acceleration, increasing the readout bandwidth from $\pm41.7$ to $\pm62.5$ kHz, and increasing the TR from 5775 to 6755 ms (which allowed for a single slice group rather than two interleaved acquisitions). The acquisition matrix of the original series was 320x256; the faster scan was able to achieve a higher matrix size of 400x280 while retaining a 21 cm field-of-view. The slice thickness was 4 mm in both cases. A triple inversion-recovery (IR) black blood cardiac protocol was optimized for shorter breath hold time and higher spatial resolution. The original protocol required 15 sec breath hold time at a resolution of 1.9x2.1 mm with 10 mm slice thickness. With AIR™ Recon DL, the breath hold time was reduced to 11 sec with a resolution of 1.2x1.4 mm and 10 mm slice thickness. Scan time reductions were achieved by increasing the parallel imaging factor from 2 to 3-fold and increasing the bandwidth from $\pm50.0$ to $\pm125.0$ kHz.

## Results

The multi-object DRO before and after passing through the AIR™ Recon DL CNN is shown in Figure 3. The AIR™ Recon DL image has visibly lower noise throughout and reduced ringing near the edges of the disks, which is most noticeable in and around the high CNR disks. The human reader study indicated that AIR™ Recon DL improved subjective low-contrast detectability in all but the single pixel disks, where detectability was equivalent to the original image. Visual detectability of the single point disks was consistent with the Rose criterion: a CNR of 5 was required to reliably detect this disk in both the original and in AIR™ Recon DL (data not shown).

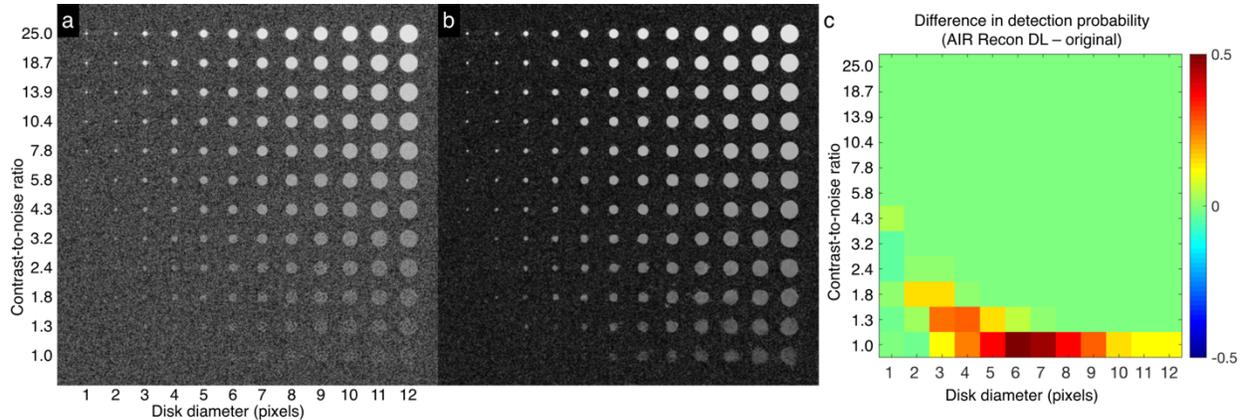

*Figure 3: Multi-object DRO and results of the human reader detectability study. One noise realization of the DRO before (a) and after (b) the AIR™ Recon DL CNN with denoising level of 0.75. The difference in detection probability (c), as assessed by human readers, indicates that AIR™ Recon DL improves detectability in all but the smallest disc, where it was nearly unchanged. Images in panels (a, b) are presented in a logarithmic intensity scale to capture the full range of signal intensities in the object.*

Results of the model observer test on the DROs is shown in Figure 4. Images passed through the CNN used in the AIR™ Recon DL pipeline with a denoising level of 0.75 were found to have improved objective detectability over the original images, irrespective of the object size. In all cases, the AUC was significantly higher with AIR™ Recon DL; p-values from a paired t-test were below $10^{-5}$ in all cases.

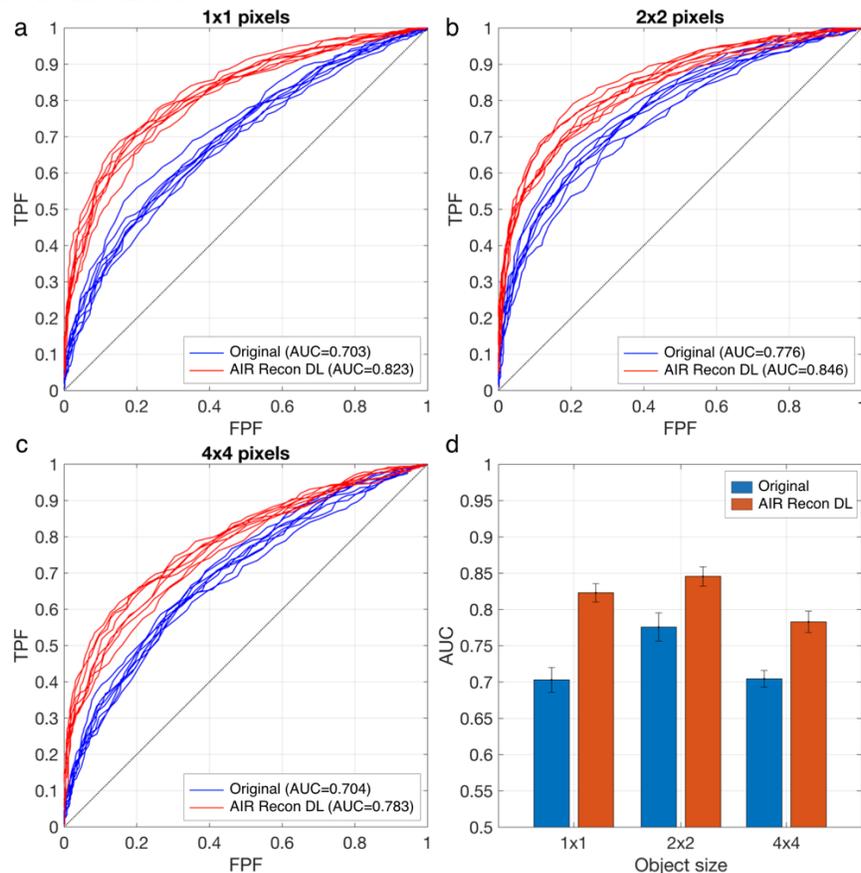

*Figure 4: ROC curves and AUC values from a signal-known-exactly model observer of original and AIR™ Recon DL images. The AUC values of AIR™ Recon DL were all statistically significant, with p<0.00001.*

Signal-to-noise ratio measurements in the ACR phantom as functions of the number of averages are shown in Figure 5. Measurements from the original image and those output from the CNN with denoising levels of 0.25, 0.5, and 0.75 were plotted. Best fit lines based on $\sqrt{\text{averages}}$ are shown and confirm that this expected relationship is observed with the original images and is maintained with the AIR™ Recon DL images. In all cases, AIR™ Recon DL improved the SNR relative to the original image; furthermore, the improvement was nearly equal to the stated denoising level.

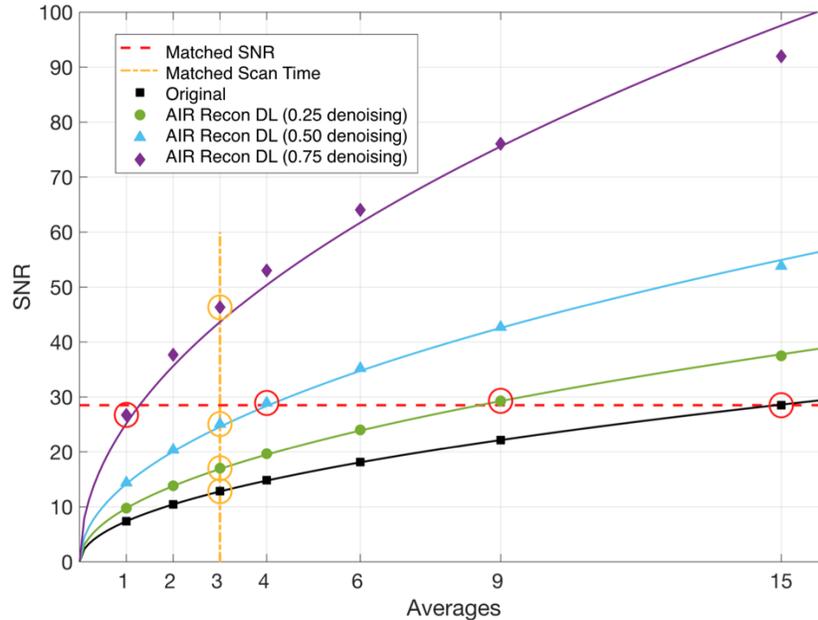

*Figure 5: Measured SNR as a function of number of averages with the original reconstruction and with various AIR™ Recon DL noise reduction settings. The best fit curves are given by $\alpha\sqrt{\text{averages}}$ where $\alpha$ is a scale factor to fit each data series. The orange vertical dashed line indicates SNR improvements at a constant scan time; the red horizontal dashed line indicates scan time reductions for near constant SNR. Images corresponding to the circled data points are shown in Figure 6.*

Two use cases for AIR™ Recon DL are illustrated by the dashed lines in Figure 5. First, an improvement in SNR with a fixed scan time is indicated by the vertical orange line. The four data points circled in orange correspond to the images shown in the top portion of Figure 6. These four images are the original, with ringing visible at the edges, and three AIR™ Recon DL images with improved SNR and no visible ringing. Second, the horizontal red line indicates a near constant SNR achieved with AIR™ Recon DL while reducing averages from 15 down to 9, 4, and 1. The data points circled in red correspond on the images in the bottom portion of Figure 6. The SNR in these four images is nearly identical despite a reduction in scan time from 97 sec in the original image down to 58 sec, 25 sec, and 6.5 sec in images with 9, 4, and 1 average(s) respectively. Unlike the original, the AIR™ Recon DL images have no visible ringing near edges. These use cases provide insights into how AIR™ Recon DL can be leveraged to adjust the number of averages in a scan protocol either to improve SNR for a given scan time, or reduce scan time while maintaining a target SNR.

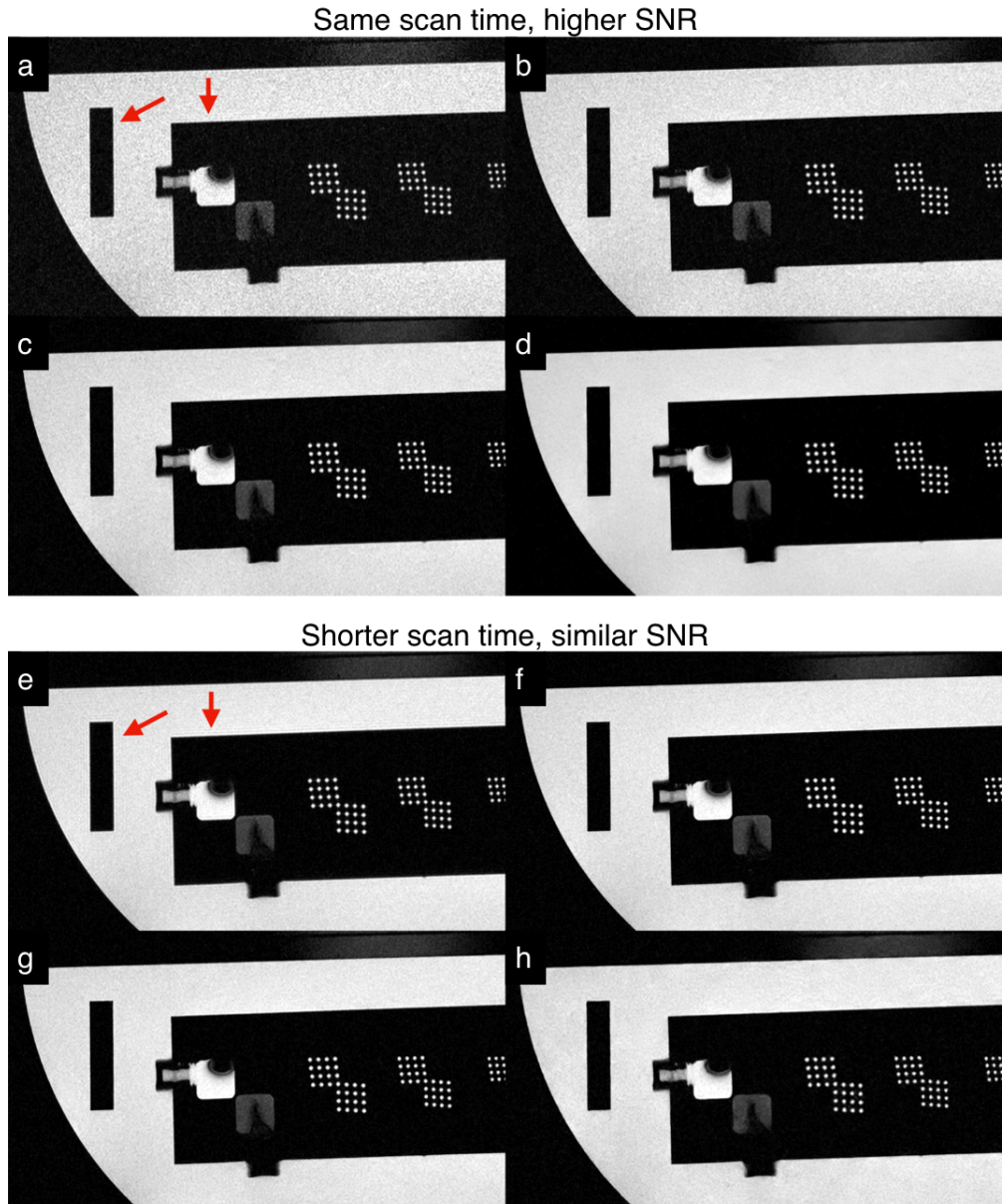

*Figure 6: Cropped portions of the ACR phantom with conventional reconstruction (a, e) and with AIR™ Recon DL at denoising levels of 0.25 (b, f), 0.50 (c, g), and 0.75 (d, h). SNR can be improved in the same scan time (a-d) or scan time can be drastically reduced from 97 sec. (e) to 58 sec. (f), 25 sec. (g) or 6.5 sec. (h) with a near-constant SNR. In all cases AIR™ Recon DL reduces truncation artifacts near edges, red arrows, and preserves image detail.*

Relative image sharpness in the ACR phantom is quantified in Figure 7. The maximum edge gradient of AIR™ Recon DL was consistently 1.6 times greater than in the original image. This improved sharpness was constant over the range of input SNR values tested and was independent of the denoising level. It was also consistent between the horizontal and vertical edges.

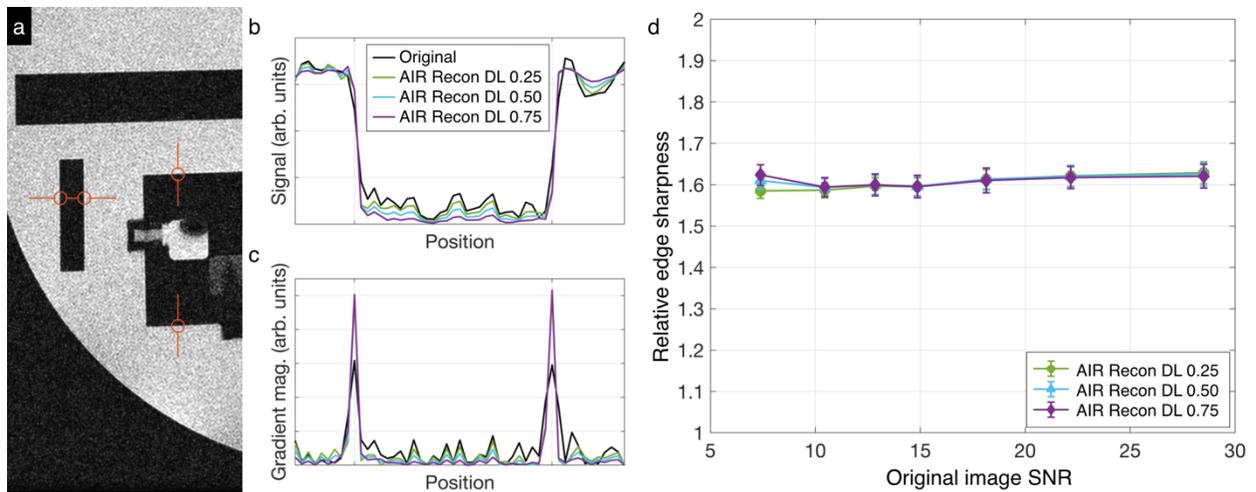

*Figure 7: Relative edge sharpness across line profiles with AIR™ Recon DL at various denoising levels. The location of 4 edges used to compute sharpness (a), representative profiles (b), and profile gradients (c) through the horizonal line. The relative edge sharpness (d) was computed from the ratio of the gradient peaks measured with AIR™ Recon DL and with the original. Edges of this phantom were 60% sharper with AIR™ Recon DL than the original, independent of the denoising level, input SNR in the phantom, and edge direction.*

Typical 3.0 T sagittal spine images with different contrast weightings are shown in Figure 8. Ringing and noise are visible in the images reconstructed with the conventional pipeline. Ringing is particularly problematic in the spinal cord, where it may mimic the central canal or nerve branches. Images reconstructed with AIR™ Recon DL show no truncation artifact and improved sharpness over the conventional images. This is particularly visible in the sub-cutaneous fat but also in the nerve roots branching out of the cord. Pathology and anatomy visible in the conventional images, such as the hyperintense $T_2$ cord lesion, vertebral heterogeneity, and bulging discs, have a higher CNR with AIR™ Recon DL making these features more evident.

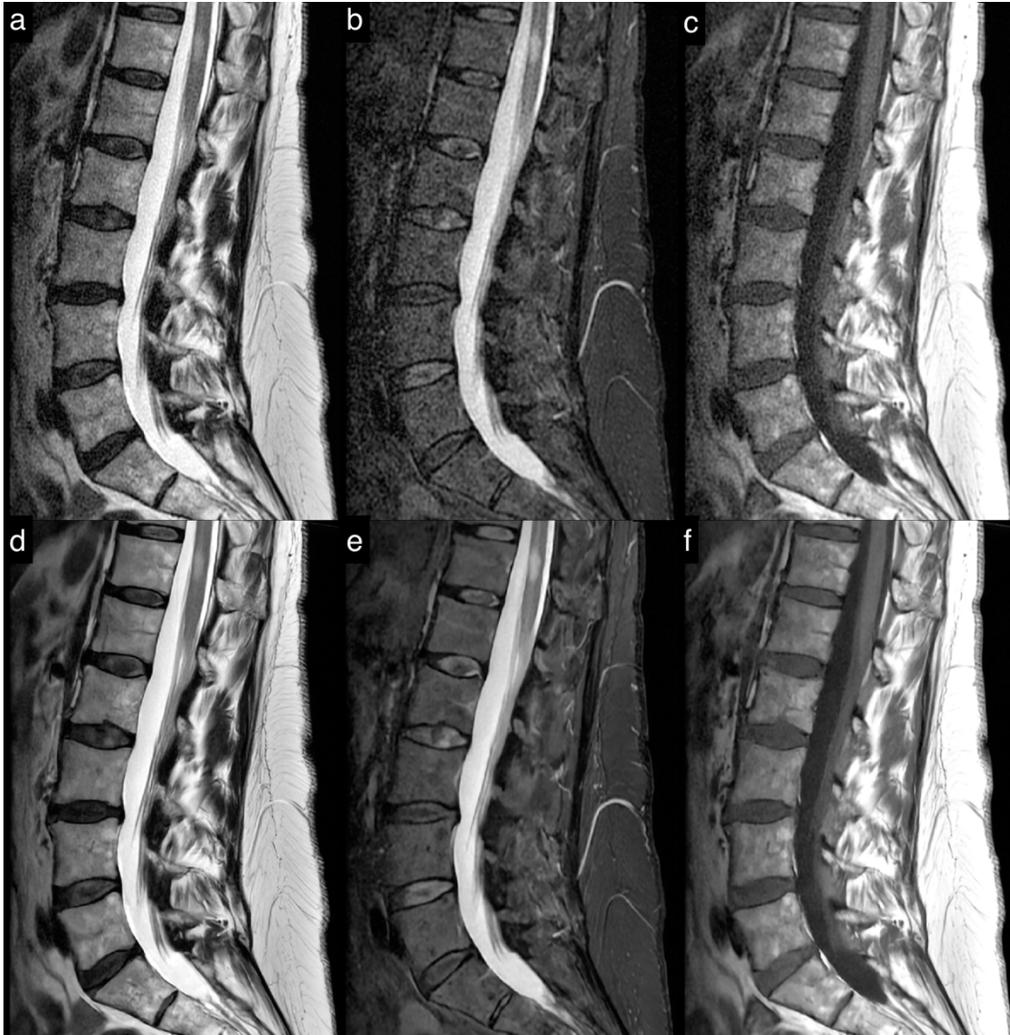

*Figure 8: Sagittal lumbar spine images reconstructed using a conventional pipeline (a–c) and with AIR™ Recon DL at a noise reduction factor of 0.75 (d–f). AIR™ Recon DL is compatible with multiple contrast weightings, including $T_2$ (a, d), $T_2$-STIR (b, e), and $T_1$ (c, f). Ringing and noise reduction are evident with AIR™ Recon DL where cord lesions and nerve branches are unobscured and contrast in the vertebrae and disk are more clearly depicted.*

Generalizability of AIR™ Recon DL to multiple anatomies, contrast weightings, pulse sequences, field strengths, and exogenous contrast agents is shown qualitatively in Figure 9. AIR™ Recon DL demonstrates visibly reduced noise levels and sharper images in all cases. The 3.0 T fat saturated wrist images (Figure 9 a, e) demonstrate that AIR™ Recon DL improves sharpness, particularly at the chondral and osseous borders, while depicting a central tear of the triangulofibral cartilage complex.

The 1.5 T coronal single shot fast spin echo abdominal images (Figure 9 b, f) demonstrate that AIR™ Recon DL is compatible with simultaneous partial Fourier and parallel imaging and enables rapid, motion-insensitive imaging with the sharpness and quality typical of multi-shot fast spin echo. With reduced noise, diffuse liver pathology is more evident in the AIR™ Recon DL image than in the conventional image. Improved sharpness is seen throughout the multiple organs and sharper vessels are visible in the liver and lungs.

Conventional reconstruction of short axis 3.0 T cardiac images (Figure 9 c, g) acquired with double-IR black-blood fast spin echo and a parallel imaging acceleration factor of 3 are noisy and poorly depict key anatomical structures. Spatially variable g-factor noise from parallel imaging is visible in the middle of the image, obscuring the myocardium. The corresponding AIR™ Recon DL image provides improved depiction of the blood-myocardial border, papillary muscles, and pericardium. Reduced noise and ringing are visible within the myocardium with AIR™ Recon DL relative to the conventional reconstruction.

Coronal contrast-enhanced brain images at 3.0 T of a possible pituitary adenoma (Figure 9 d, h) demonstrate how AIR™ Recon DL reduces noise throughout the brain while providing sharp boundaries between gray/white matter and cerebrospinal fluid and vessels and visualization of small anatomical details around the sella. Notably, AIR™ Recon DL maintains texture within the lesion and provides clear boundaries between the tumor and the adjacent carotid arteries and nearby optic nerves.

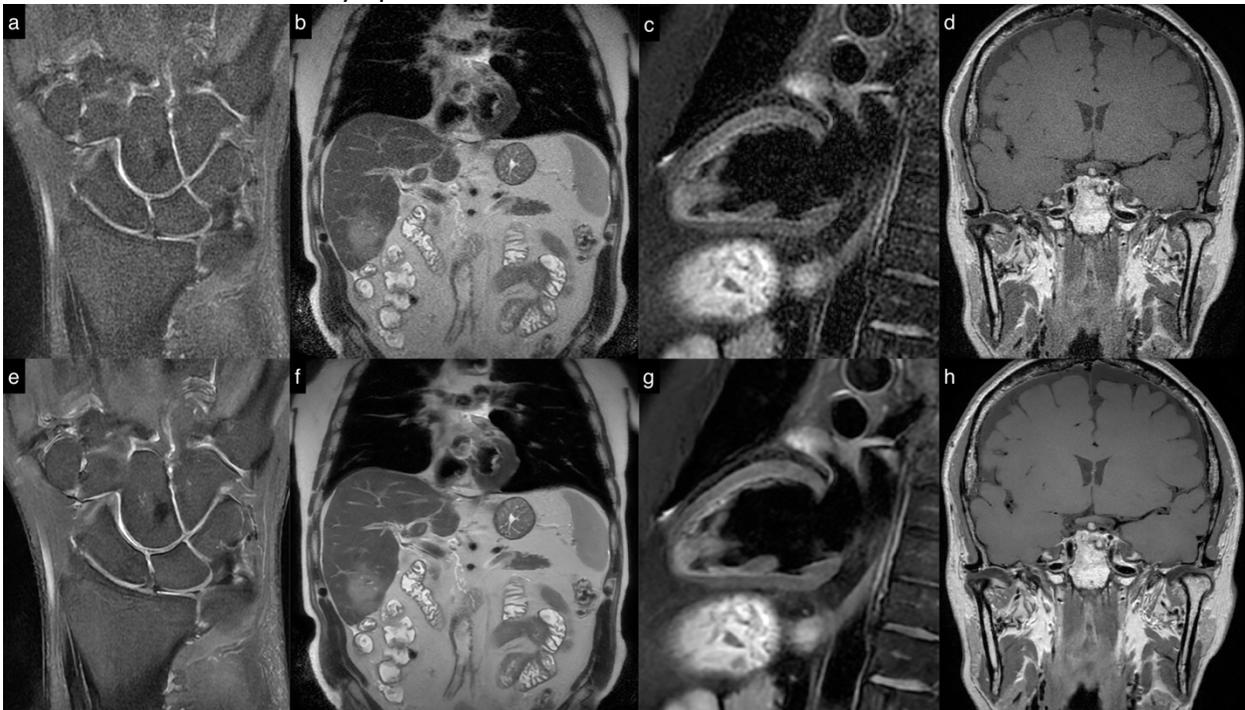

*Figure 9: Images from diverse contrast weightings, pulse sequences, and anatomies reconstructed with a conventional pipeline (a–d) and with AIR™ Recon DL with a 0.75 noise reduction factor (e–h). a, e: Fat sat T2 weighted fast spin echo of the wrist at 3.0 T. b, f: $T_2$ weighted single shot fast spin echo of the abdomen at 1.5 T. c, g: Double-IR fast spin echo of the heart at 3.0 T. d, h: Contrast enhanced $T_1$ weighted fast spin echo of the brain at 3.0 T.*

The SNR and sharpness improvements afforded by AIR™ Recon DL were leveraged to optimize protocols for the purpose of reduced scan time. Figure 10 shows images from two standard clinical protocols (a, b) and shorter variants with near-identical contrast weighting (c, d). The fast AIR™ Recon DL $T_2$ weighted image (c) is visually sharper than the longer conventional image (a), with a well-defined midline fissure providing clear hemispheric separation and a continuous lining of cerebrospinal fluid visible between the temporal lobes and the cerebellum. The accelerated $T_2$ protocol was acquired in 61 sec while the original protocol required 127 sec. Time savings were achieved by a combination of increased parallel imaging factor (from 1.0x to 1.5x), higher readout bandwidth, and fewer slice groupings. Improved resolution in the cardiac protocol provides clear visualization of the myocardium and pericardium while reduced noise in the blood pool improves image contrast. With AIR™ Recon DL the echo spacing could be reduced from 5.5 to 4.3 ms leading to a temporal footprint reduction from 102 to 93 ms, which is expected to reduce motion artifacts. Fewer segments with the modified protocol yield a breath hold reduction from 15 to 11 sec, which will typically improve patient compliance, reduce exam times, and improve image quality.

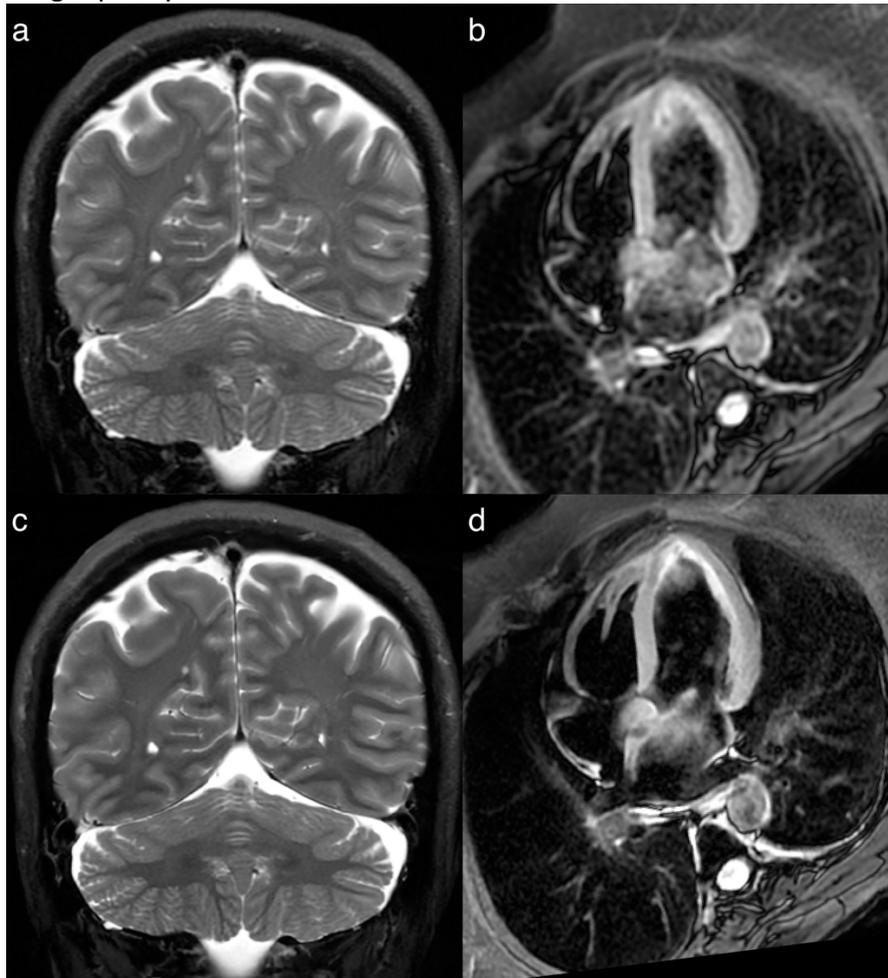

*Figure 10: The increased SNR and sharpness afforded by AIR™ Recon DL enable protocol optimizations to reduce scan time and increase matrix size. Original $T_2$ weighted brain (a) and triple-IR black-blood cardiac (b) image series were acquired in 127 sec and 15 sec respectively. Increased readout bandwidth, timing optimizations, and parallel imaging enable scan times of 61 sec and 11 sec for similar $T_2$ (c) and black-blood (d) sequences respectively, reconstructed with AIR™ Recon DL. Slight positioning and contrast changes are due to motion between series and minor timing differences.*

## Discussion

AIR™ Recon DL is an image reconstruction pipeline that includes a deep CNN to overcome limitations of conventional reconstruction pipelines, namely ineffective denoising, image blurring, and residual ringing. The aim is to maximize relevant information and minimize distractors. AIR™ Recon DL retains the effective and well understood elements of the image reconstruction pipeline, including the inverse Fourier transform and acceleration via ARC (Brau, Beatty, Skare, & Bammer, 2006), ASSET (King, 2004), and partial Fourier, making it broadly compatible with existing applications and protocols. AIR™ Recon DL removes the previous filters used to denoise and/or mitigate ringing. These filters blur the image and only partially remove ringing. AIR™ Recon DL was designed to consistently suppress this ringing artifact without reducing resolution while allowing the user to adjust the denoising level based on individual preference. Recently, another group developed a clever solution to solve a similar problem (Muckley et al., 2019). This solution employed natural images to train their network and demonstrated excellent performance for enhancing diffusion images. This work was restricted improving diffusion DICOM images (although could presumably by generalized to other use cases) but does not provide tunable denoising. The objectives of AIR™ Recon DL are similar to this work; however, our solution is focused on generalizability and is tightly integrated with the reconstruction process.

The CNN is integrated inside the reconstruction pipeline and operates on raw, complex-valued data. This is important since the CNN occurs prior to operators that irreparably alter the signal and noise characteristics. An absolute value operation is commonly used to remove image phase and display the image magnitude, but this rectifies noise and introduces a signal-dependent bias that is very difficult to reverse. Gradient unwarping expands or compresses regions of the image using an interpolation kernel to preserve accurate spatial positioning, but this alters the direction and spatial frequency of the truncation artifact and introduces spatially variable noise correlations. Surface coil intensity correction scales the signal and noise in a spatially dependent manner. DICOM export involves integer truncation, which reduces the signal bit depth and imposes a scale dependence on subsequent image manipulations. Ultimately, post-processing image enhancements are limited in their ability to improve the image quality and a reconstruction integrated solution is preferred.

AIR™ Recon DL was found to provide up to 60% sharper edges in the ACR phantom (Figure 7), relative to a raw image. This is an ideal test case since it contains abrupt edges without partial volume effects and minimal confounding structures. Improved sharpness is observed *in vivo* but a 60% improvement is likely an upper limit. Improved edge sharpness is achieved by leveraging truncation artifact as an indicator of missing information, rather than simply an artifact to be suppressed. AIR™ Recon DL provides effective interpolation by estimating the high spatial frequency information needed to support the acquired data. Conventional pipelines do the opposite: portions of the acquired k-space data are attenuated for consistency with the unsampled high frequency data, ultimately blurring the image.

The supervised learning approach that employed a diverse collection of source images and image augmentations enabled broad generalizability of AIR™ Recon DL to multiple anatomies, pulse sequences, contrast weightings, and field strengths. Furthermore, it is compatible with any coil configuration, including single channel and quadrature coils. The reconstruction pipeline is effectively agnostic to the input data. Figure 6 shows typical performance on a phantom using a quadrature head coil; Figure 8 shows a diversity of contrast weightings in the spine with a 32-channel posterior/anterior array; Figure 9 shows multiple anatomies, pulse sequence families, scan planes, contrast weightings, and field strengths.

Balancing scan time, SNR, and resolution is a perpetual compromise with MR and few protocols achieve the desired balance. The SNR and sharpness benefit of AIR™ Recon DL can be leveraged to improve the image quality of existing protocols, as shown in Figure 8 and Figure 9.

The benefits of AIR™ Recon DL can be leveraged to optimize protocols with more flexibility. Common strategies for reducing scan time involve decreasing the number of averages, increasing bandwidth, and/or increasing the parallel imaging acceleration factor. Figure 10 shows two examples of substantially faster protocols where both the SNR and sharpness appear to be improved in the faster scan reconstructed with AIR™ Recon DL. These examples also demonstrate how the acquisition matrix can be increased while simultaneously reducing scan time.

AIR™ Recon DL was found to improve low-contrast detectability in two DROs. The worst performance was in single point objects assessed by human readers where detectability was similar to the original images. In this case, detectability was governed by the Rose criteria, where a contrast-to-noise-ratio of 5 is required for reliable detection. Single point objects are generally the most difficult to detect since they have no spatial correlations. At very low SNR levels, where the signal is fully lost in the noise, the low contrast detectability of AIR™ Recon DL is similar to the conventional reconstruction.

AIR™ Recon DL was not designed to remove or alter the appearance of other common artifacts, like motion, flow, balanced SSFB banding, and ghosting. Coherent signals, desirable or not, that are encoded in the data are retained in the image, Figure 2. The appearance of these artifacts is largely unchanged and so remain familiar and distinct from pathology, to the clinician. Motion related artifacts might be addressed indirectly with shorter scan times enabled by AIR™ Recon DL.

## Conclusions

AIR™ Recon DL is an image reconstruction pipeline leveraging artificial intelligence that is shown through phantom and in vivo results to improve image SNR and sharpness while reducing truncation artifacts. It is broadly compatible across multiple anatomies, contrast weightings, and imaging situations. It can be applied to improve the image quality of existing protocols and/or leveraged to optimize protocols for reduced scan time, offering a powerful new tool to improve the diagnostic and operational performance of an MRI scanner.